\documentclass[final,3p,times]{elsarticle}

\usepackage{amssymb}

\journal{Physic Procedia}

\usepackage{graphicx}
\usepackage{epstopdf}
\usepackage{hyperref}

\begin{document}

\begin{frontmatter}



\title{BaFe$_2$As$_2$/Fe bilayers with [001]-tilt grain boundary on MgO and SrTiO$_3$ bicrystal substrates}

\author[label1]{Kazumasa\,Iida\corref{cor1}}
\ead{k.iida@ifw-dresden.de}
\author{Silvia\,Haindl\fnref{label2}}
\author{Fritz\,Kurth\fnref{label1}}
\author{Jens\,H\"{a}nisch\fnref{label1}}
\author{Ludwig\,Schultz\fnref{label1}}
\author{Bernhard\,Holzapfel\fnref{label1}}

\address[label1]{Institute for Metallic Materials, IFW Dresden. 01171 Dresden, Germany}
\address[label2]{Institute for Solid State Research, IFW Dresden. 01171 Dresden, Germany}

\cortext[cor1]{Corresponding author: Kazumasa\,Iida; Tel.: +49-351-4659-608; Fax: +49-351-4659-541.}

\begin{abstract}
Co-doped BaFe$_2$As$_2$ (Ba-122) can be realized on both MgO and SrTiO$_3$ bicrystal substrates with [001]-tilt grain boundary by employing Fe buffer layers. However, an additional spinel (i.e. MgAl$_2$O$_4$) buffer between Fe and SrTiO$_3$ is necessary since an epitaxial, smooth surface of Fe layer can not be grown on bare SrTiO$_3$. Both types of bicrystal films show good crystalline quality.
\end{abstract}


\end{frontmatter}


\section{Introduction}
In high-$T_{\rm c}$ cuprates, [001]-tilt grain boundaries (GBs) in MgO and SrTiO$_3$ bicrystal substrates have been commonly used for realizing Josephson junctions as well as investigating transport properties of GBs. After the discovery of Fe-based pnictide superconductors,\cite{1} the community became immediately interested in exploring grain boundaries in this new class of superconductors. 

Despite the recent success of realizing Co-doped BaFe$_2$As$_2$ (Ba-122) on [001]-tilt GB MgO and SrTiO$_3$ bicrystal substrates,\cite{2,3} these substrates may not be suitable for several reasons. Firstly, a large misfit between Ba-122 and MgO of around 6\% leads to low crystalline quality or even non-epitaxial growth under our deposition condition.\cite{4} Secondly, the SrTiO$_3$ substrate becomes electrically conductive under ultra-high vacuum (UHV) conditions, which may compromise transport measurements particularly near the normal-superconducting transition. However, implementation of Fe for MgO and Fe+MgAl$_2$O$_4$ for SrTiO$_3$ as buffer layers can solve those problems. An Fe layer itself is electrically conductive, however, the recalculation of the superconducting transition temperature ($T_{\rm c}$) and the critical current density ($J_{\rm c}$) is possible due to its thin layer thickness (20\,nm).\cite{5} In this report, the fabrication of Co-doped Ba-122 thin films on both [001]-tilt bicrystal MgO and SrTiO$_3$ substrates and their transport properties are presented in detail. For the aim of bicrystal Josephson junctions, substrates with a large [001]-tilt angle of over 20$^\circ$ have been used in this investigation.

\section{Experimental procedure}
\subsection{Ba-122 on [001]-tilt MgO bicrystal}
A 20 nm thick Fe buffer layer was deposited by means of pulse laser deposition (PLD) on a [001]-tilt MgO bicrystal substrate with misorientation angle of $\theta_{\rm GB}$=36.8$^\circ$ (nominal value) at room temperature, followed by a high-temperature annealing at 700\,$^\circ$C. This process is very effective for obtaining epitaxial Fe layers with flat surface. After the Fe buffer preparation, Co-doped Ba-122 (BaFe$_{1.8}$Co$_{0.2}$As$_2$) with 100\,nm thickness was deposited at 700\,$^\circ$C. All deposition processes have been conducted under UHV condition. A detailed description of both Fe buffer and Co-doped Ba-122 deposition procedure can be found in Ref.\cite{6}. After the structural characterization described in {\it sub-section 2.3}, Au layers were deposited on the films by PLD at room temperature followed by ion beam etching to fabricate bridges of 0.5\,mm width and 1\,mm length for transport measurements.

\subsection{Ba-122 on [001]-tilt SrTiO$_3$ bicrystal}
A MgAl$_2$O$_4$ buffer was deposited on a [001]-tilt GB SrTiO$_3$ bicrystal substrate with $\theta_{\rm GB}$=30$^\circ$ at 650\,$^\circ$C by PLD in UHV condition (pulse number 4000 at 40\,mJ). In a pilot experiment, we have confirmed that MgAl$_2$O$_4$ can be grown epitaxially on SrTiO$_3$ (100) substrate. After the primary buffer deposition, the MgAl$_2$O$_4$-buffered SrTiO$_3$ substrate was cooled to room temperature for the secondary, Fe buffer deposition. A 20\,nm thick Fe buffer was prepared by the same procedure as described in {\it sub-section 2.1} except for the annealing temperature. Here, a slightly higher temperature of 750\,$^\circ$C was employed. After the Fe buffer preparation, a 100\,nm thick Co-doped Ba-122 (BaFe$_{1.84}$Co$_{0.16}$As$_2$) layer was deposited. {\it In-situ} gold layer deposition at room temperature was conducted after the Ba-122 deposition for the aim of protection from possible damage during the structuring process.

\subsection{Structural characterization and transport measurement}
Phase purity and out-of-plane texture were investigated by means of X-ray diffraction in Bragg-Brentano geometry with Co-K$_\alpha$ radiation. In-plane orientation of both Fe and Co-doped Ba-122 were investigated by using the 110 and 103 poles, respectively, in a texture goniometer operating with Cu-K$_\alpha$ radiation. The respective measured reflection of the MgO and SrTiO$_3$ bicrystal substrates for in-plane texture measurements are 220 and 110.

Superconducting properties were measured in a Physical Property Measurement System (PPMS, Quantum Design) by a standard four-probe method with a criterion of 1\,$\rm\mu Vcm^{-1}$ for evaluating $J_{\rm c}$.

\section{Results and discussion}
\begin{figure}[b]
\centering
		\includegraphics[width=130mm]{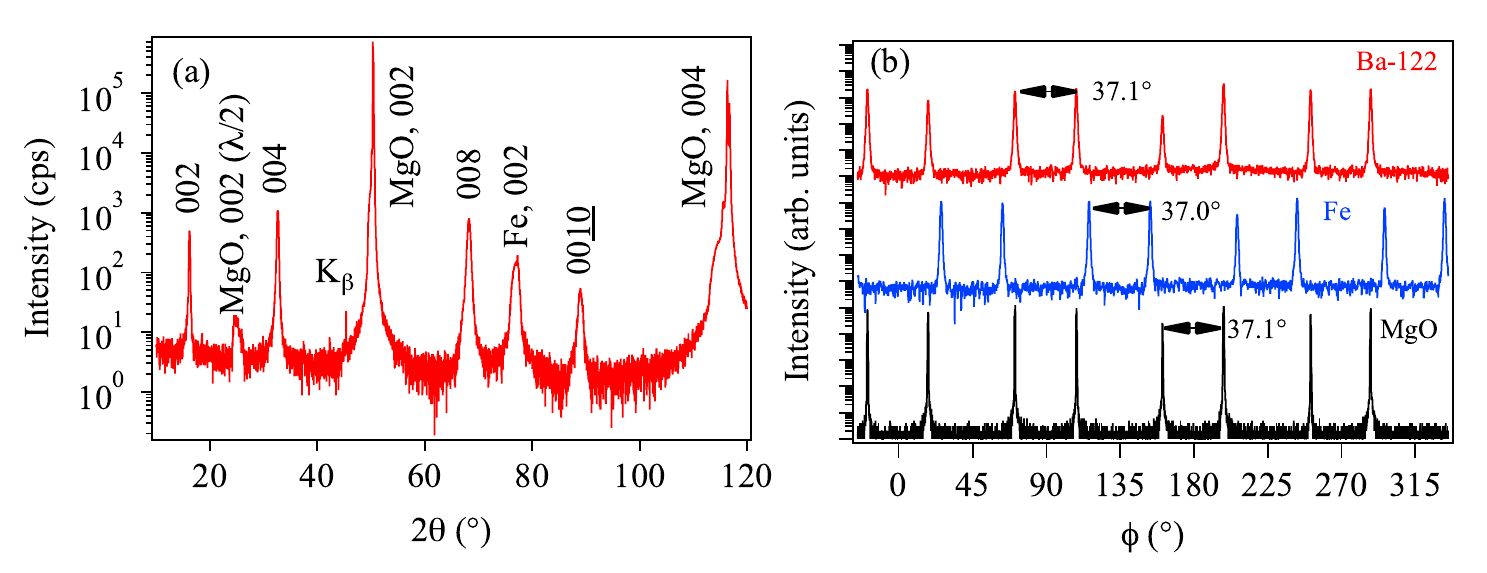}
		\caption{(a) The $\theta/2\theta$ scan of Ba-122/Fe bilayer on [001]-tilt GB MgO bicrystal substrate in Bragg-Brentano geometry using Co-K$_\alpha$ radiation. (b) The $\phi$ scans of the 103 Ba-122, the 110 Fe and the 220 MgO. Average misorientation angles are 37.1$^\circ$, 37.0$^\circ$, and 37.1$^\circ$ for Ba-122, Fe and MgO, respectively.} 
\label{fig:figure1}
\end{figure}

XRD patterns in logarithmic scale (fig.\,\ref{fig:figure1}(a)) clearly show that the film has been grown with high phase purity and $c$-axis texture, i.e. with [001] perpendicular to the substrate surface. Fig.\,\ref{fig:figure1}(b) shows the $\phi$ scans of the 103 Ba-122, the 110 Fe and the 220 MgO. It is evident from fig.\,\ref{fig:figure1}(b) that two grains with $\theta_{\rm GB}\sim37.1^\circ$, which is slightly larger than the nominal value of 36.8$^\circ$, are grown due to the perfect transfer of the grain orientation from the substrate via Fe buffer to the Ba-122 layer. Here the respective average $\Delta\phi_{\rm Ba-122}$ and $\Delta\phi_{\rm Fe}$ are 0.84$^\circ$ and 0.80$^\circ$ (not corrected for instrumental broadening), indicative of good in-plane alignment for both layers. The GB is symmetric, i.e. it is situated with $\theta_{\rm GB}/2$ to (100) in both grains.

\begin{figure}[t]
\centering
		\includegraphics[width=160mm]{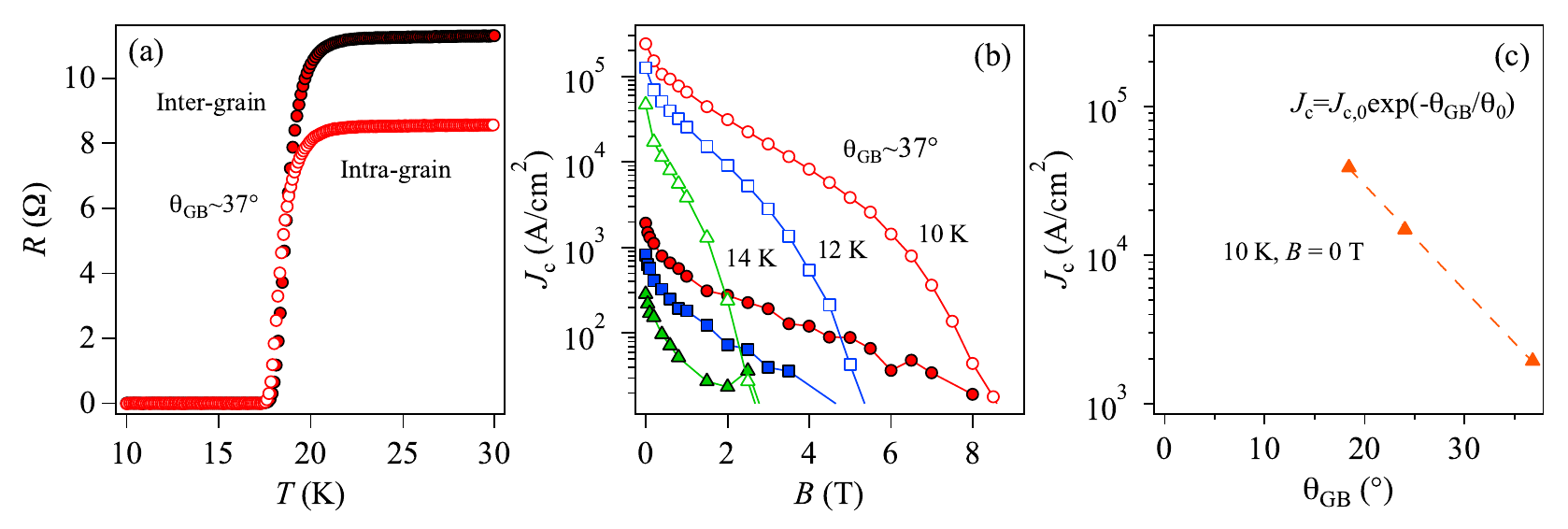}
		\caption{(a) $R-T$ curves of inter- and intra-grains. Both curves show the same $T_{\rm c}$ of around 20\,K. (b) $J_{\rm c}-B$ performances ($B\parallel c$) of the corresponding grains at different temperatures. Open symbols represent the intra-grain and the solid symbols show the inter-grain curves. (c) Normalized critical current density ($J_{\rm c}^{\rm inter}/J_{\rm c}^{\rm intra}$) as a function of $\theta_{\rm GB}$ at 10\,K.} 
\label{fig:figure2}
\end{figure}

$R-T$ curves in fig.\,\ref{fig:figure2}(a) show that both the inter- and intra-grain bridge have the same onset $T_{\rm c}$ as well as zero resistance, indicating the high quality of the bicrystal films. The field dependence of $J_{\rm c}$ ($B\parallel c$) for both the inter- and intra-grain bridge at various temperatures is displayed in fig.\,\ref{fig:figure2}(b). Clearly, the intra-grain $J_{\rm c}$ ($J_{\rm c}^{\rm intra}$) is higher than the inter-grain $J_{\rm c}$ ($J_{\rm c}^{\rm inter}$) at low magnetic fields regime due to the large $\theta_{\rm GB}$. The difference between $J_{\rm c}^{\rm intra}$ and $J_{\rm c}^{\rm inter}$ is getting smaller with increasing applied field and finally both curves overlap at the irreversibility field for $\theta_{\rm GB}\sim37^\circ$. On the assumption that the weak-link behavior is empirically described by $J_{\rm c}^{\rm inter}=J_{\rm c}^{\rm intra}{\rm exp}(-\theta_{\rm GB}/\theta_{\rm 0})$ in low field regime, the characteristic angle ($\theta_{\rm 0}$) is estimated to around 7.9$^\circ$ for our Co-doped Ba-122 bicrystal films (fig.\,\ref{fig:figure2}(c)), which is almost twice as large as that of YBa$_2$Cu$_3$O$_7$.\cite{7}

\begin{figure}[b]
\centering
		\includegraphics[width=120mm]{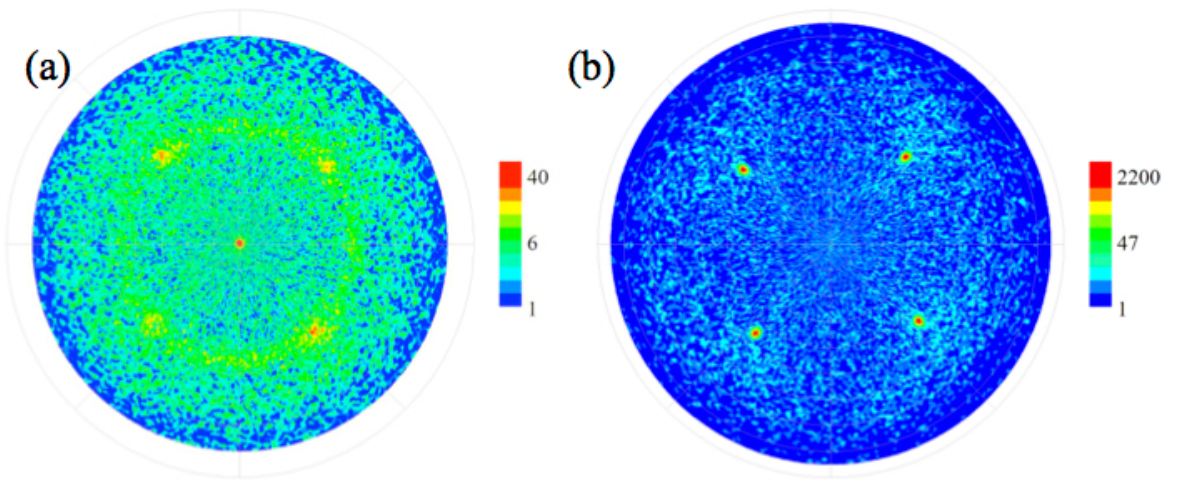}
		\caption{(a) The 110 pole figure measurement of Fe on bare SrTiO$_3$ and (b) MgAl$_2$O$_4$-buffered SrTiO$_3$.} 
\label{fig:figure3}
\end{figure}

The growth of epitaxial Fe with smooth surface on SrTiO$_3$ substrate is not straightforward, which is presumably due to the difference in crystal structure rather than to the lattice mismatch. As a result, epitaxial growth of Ba-122 is hardly achieved on Fe-buffered SrTiO$_3$. Indeed, fiber textured Fe is observed (fig.\,\ref{fig:figure3}(a)), when it is prepared directly on SrTiO$_3$ substrate with the same procedure as on MgO substrates. In order to avoid this problem, an additional buffer layer of MgAl$_2$O$_4$ between Fe and SrTiO$_3$ substrate has been used. MgAl$_2$O$_4$ fits better to accommodate the lattice mismatch to Fe rather than MgO as shown in Table\,\ref{tab:table1}. Here, the lattice mismatch is defined as $(a_{\rm f}-a_{\rm s})/a_{\rm f}$, where $a_{\rm f}$ and $a_{\rm s}$ are the lattice parameters of Fe and substrates, respectively. Additionally, a small lattice mismatch of 3.4\% between MgAl$_2$O$_4$ and SrTiO$_3$ may lead to good epitaxial growth of the MgAl$_2$O$_4$ buffer.

\begin{table}
\centering
\caption{\label{tab:table1}The crystal structure, lattice parameter and the lattice mismatch to Fe of MgO and MgAl$_2$O$_4$.}
\small
\begin{tabular}{cccc}\hline
Materials&Crystal structure&Lattice parameter\,(nm)&Lattice mismatch\,(\%)\\\hline
MgO&Rock salt&0.421&3.7\\
MgAl$_2$O$_4$&Spinel&0.8083 ($a$/2=0.4042)&-0.4\\\hline
\end{tabular}
\end{table}

Fig.\,\ref{fig:figure3}(b) shows the 110 pole figure of Fe on MgAl$_2$O$_4$-buffered SrTiO$_3$ fabricated by the same procedure described in {\it sub-section 2.1}. It is clear from fig.\,\ref{fig:figure3}(b) that a perfect textured growth of Fe is realized (average $\Delta\phi_{\rm Fe}=0.61^\circ$). During heating of Fe, evolution of the Fe-layer texture was observed through reflection high energy electron diffraction (RHEED), which is a similar observation of Fe on single crystalline MgO substrates.\cite{8} The RHEED images of the Fe layer on MgAl$_2$O$_4$-buffered SrTiO$_3$ after the high temperature annealing showed only streak patterns, indicative of surface smoothing. This architecture also prevents the current-shunting effect between SrTiO$_3$ and Co-doped Ba-122. This buffer architecture offers the opportunity to grow smooth epitaxial Fe layers on various perovskite substrate such as (La,Sr)(Al,Ta)O$_3$ (100) substrate (not shown in this paper).
 
\begin{figure}
\centering
		\includegraphics[width=160mm]{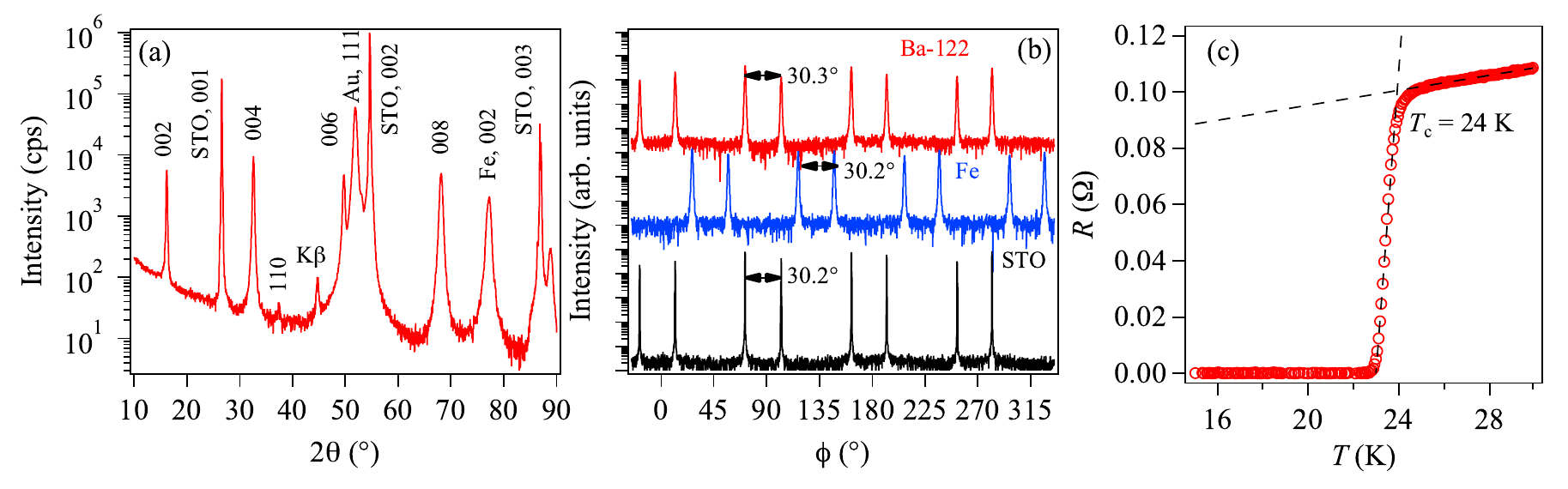}
		\caption{(a) The $\theta/2\theta$ scan of Ba-122/Fe/MgAl$_2$O$_4$ trilayer on [001]-tilt SrTiO$_3$ (STO) bicrystal substrate in Bragg- Brentano geometry using Co-K$_\alpha$ radiation. (b) The $\phi$ scans of the 103 Ba-122, the 110 Fe, and 110 STO. Average misorientation angles are 30.3$^\circ$, 30.2$^\circ$, and 30.2$^\circ$ for Ba-122, Fe and STO, respectively. (c) Intra-grain $T_{\rm c}$ was recorded as 24\,K.} 
\label{fig:figure4}
\end{figure}
 
Structural characterization of Ba-122/Fe/MgAl$_2$O$_4$ on [001]-tilt GB SrTiO$_3$ substrate by means of X-ray diffraction is summarized in figs.\,\ref{fig:figure4}. As stated earlier, {\it in-situ} Au layer deposition was conducted after the Ba-122 deposition, which explains the 111 reflection of Au in fig.\,\ref{fig:figure4}(a). It is further clear from fig.\,\ref{fig:figure4}(a) that only 00$l$ reflections of Ba-122 with a small amount of the 110 reflection are observed. Fig.\,\ref{fig:figure4}(b) displays the $\phi$ scans of the 103 Ba-122, the 110 Fe and the 110 SrTiO$_3$. Clear 8 peaks from two grains with a rotational angle of 30$^\circ$ are apparent. Accordingly, an epitaxial Ba-122 film is formed on [001]-tilt SrTiO$_3$ bicrystal substrate. The onset $T_{\rm c}$=24\,K of the film is almost identical to that of Ba-122/Fe bilayers on single crystal substrates (fig.\,\ref{fig:figure4}(c)). This value is higher than the intra-grain $T_{\rm c}$ deposited on MgO bicrystal presented in fig.\,\ref{fig:figure2}(a) due to a higher deposition temperature as well as different PLD targets (i.e. different Co concentration). The resultant bicrystal junctions show clear Josephson effects and the detailed studies can be found in Ref.\cite{9}.

\section{Summary}
High-quality Co-doped Ba-122 bicrystal films can be realized on both [001]-tilt MgO and SrTiO$_3$ substrates by employing Fe buffer layers via pulsed laser deposition. The additional MgAl$_2$O$_4$ buffer between Fe and SrTiO$_3$ is necessary for realizing epitaxial Ba-122. The characteristic angle of Co-doped Ba-122 in low field regime is about 8$^\circ$, which is almost twice as large as that of YBa$_2$Cu$_3$O$_7$.

\section*{Acknowledgments}
The authors would like to thank R.\,H\"{u}hne for fruitful discussions, M.\,K\"{u}hnel, and U.\,Besold for their technical support. This work was supported by DFG under Project Nos. BE\,1749/13 and HA\,5934/3-1. The research leading to these results has furthermore received funding from European Union's Seventh Framework Programe (FP7/2007-2013) under grant agreement number 283141 (IRON-SEA).


\section*{References}
\begin{thebibliography}{99}
\bibitem{1}
Y. Kamihara, T. Watanabe, M. Hirano, H. Hosono, J. Am. Chem. Soc. {\bf130} (2008) 3296.
\bibitem{2}
S. Lee, J. Jiang, J. D. Weiss, C. M. Folkman, C. W. Bark, C. Tarantini $et$ $al$., Appl. Phys. Lett. {\bf95} (2009) 212505.
\bibitem{3}
T. Katase, Y. Ishimaru, A. Tsukamoto, H. Hiramatsu, T. Kamiya, K. Tanabe, H. Hosono, Nat. Commun. {\bf2} (2011) 409.
\bibitem{4}
T. Thersleff, K. Iida, S. Haindl, M. Kidszun, D. Pohl, A. Hartmann $et$ $al$., Appl. Phys. Lett. {\bf97} (2010) 022506.
\bibitem{5}
S. Trommler, R. H{\"u}hne, J. H{\"a}nisch, E. Reich, K. Iida, S. Haindl $et$ $al$., Appl. Phys. Lett. {\bf100} (2012) 122602.
\bibitem{6}
K. Iida, S. Haindl, T. Thersleff, J. H{\"a}nisch, F. Kurth, M. Kidszun $et$ $al$., Appl. Phys. Lett. {\bf97} (2010) 172507.
\bibitem{7}
H. Hilgenkamp, J. Mannhart, Rev. Mod. Phys. {\bf74} (2002) 485.
\bibitem{8}
K. Iida, J H{\"a}nisch, S. Trommler, S. Haindl, F. Kurth, R. H{\"u}hne  $et$ $al$., Supercond. Sci. Technol. {\bf24} (2011) 125009.
\bibitem{9}
S. Schmidt, S. D{\"o}ring, F. Schmidl, V. Tympel, S. Haindl, K. Iida $et$ $al$., arXiv:1211.3879v1, Accepted in IEEE Trans. Appl. Supercond.
\end{thebibliography}
\end{document}